\documentclass{article}
\usepackage{amssymb}
\usepackage{amsfonts}
\usepackage{amsmath}
\usepackage{cite}
\allowdisplaybreaks
\numberwithin{equation}{section}
\DeclareMathOperator{\sgn}{\rm sgn}
\DeclareMathOperator{\diag}{\rm diag}
\DeclareMathOperator*{\res}{\rm res}

\DeclareMathOperator{\Ko}{\mathcal{K}}
\DeclareMathOperator{\Fo}{\mathcal{F}}
\DeclareMathOperator{\Do}{\mathcal{D}}
\DeclareMathOperator{\Vo}{\mathcal{V}}

\DeclareMathOperator{\bk}{\mathbf{k}}
\newtheorem{remark}{Remark}[section]
\begin{document}

\title{On the equivalence of different approaches for generating
multisoliton solutions of the KPII equation}
\author{M.~Boiti${}^{*}$, F.~Pempinelli${}^{*}$, A.~K.~Pogrebkov${}^{\dag}$,
and B.~Prinari${}^{\ddag *}$ \\
${}^{*}$Dipartimento di Fisica, Universit\`a del Salento and\\
Sezione INFN, Lecce, Italy\\
${}^{\dag}$Steklov Mathematical Institute, Moscow, Russia\\
${}^{\ddag}$Department of Mathematics, UCCS, Colorado Springs, CO (USA)}
\date{PACS: 02.30Ik, 02.30Jr, 05.45Yv}
\maketitle

\begin{abstract}
The unexpectedly rich structure of the multisoliton solutions of the KPII equation has been explored by using different approaches, running from dressing method to twisting transformations and to the $\tau $-function formulation. All these approaches proved to be useful in order to display different properties of these solutions and their related Jost solutions. The aim of this paper is to establish the explicit formulae relating all these approaches. In addition some hidden invariance properties of these multisoliton solutions are discussed.
\end{abstract}

\section{Introduction}

The Kadomtsev--Petviashvili (KP) equation in its version called KPII
\begin{equation}
(u_{t}-6uu_{x_{1}}+u_{x_{1}x_{1}x_{1}})_{x_{1}}=-3u_{x_{2}x_{2}},
\label{KPII}
\end{equation}
where $u=u(x,t)$, $x=(x_{1},x_{2})$ and subscripts $x_{1}$, $x_{2}$ and $t$ denote partial derivatives, is a (2+1)-dimensional generalization of the celebrated Korteweg--de~Vries (KdV) equation. There are two inequivalent versions of the KP equations, corresponding to the choice for the sign in the rhs as in~(\ref{KPII}) and to the alternative choice, which is referred to as KPI equation. The KP equations, originally derived as a model for small-amplitude, long-wavelength, weakly two-dimensional waves in a weakly dispersive medium \cite{KP1970}, were known to be integrable since the beginning of the 1970s \cite{D1974,ZS1974}, and can be considered as prototypical (2+1)-dimensional integrable equations.

The KPII equation is integrable via its association to the operator
\begin{equation}
\mathcal{L}( x,\partial_{x}) =-\partial_{x_{2}}+\partial_{x_{1}}^{2}-u(x),
\label{heatop}
\end{equation}
which defines the well known equation of heat conduction, or heat equation for short.

The spectral theory of the operator~(\ref{heatop}) was developed in~\cite{BarYacoov,Lipovsky,Wickerhauser,Grinevich0} in the case of a real potential $u(x)$ rapidly decaying at spatial infinity, which, however, is not the most interesting case, since the KPII equation was just proposed in~\cite{KP1970} in order to deal with two dimensional weak transverse perturbation of the one soliton solution of the KdV. In fact, KPII admits a one soliton solution of the form
\begin{equation}
u(x,t)=-\frac{(a-b)^{2}}{2}\text{\textrm{sech}}^{2}\left[ \frac{(a-b)}{2}
x_{1}+\frac{(a^{2}-b^{2})}{2}x_{2}-2(a^{3}-b^{3})t)\right] \,,  \label{1-sol}
\end{equation}
where $a$ and $b$ are real, arbitrary constants. Multisoliton solutions have also been obtained by different methods, in \cite{Satsuma} trough Hirota method, in~\cite{MZBIM1977} and~\cite{Konop1993} via dressing, in~\cite{FN} by the Wronskian technique and in~\cite{MS1991} by using Darboux transformations. In addition, in contrast with the KPI equation, also non elastic and resonant scattering of solitons was described in \cite{M1977,Z1986, KPIIproceedings, Pr2000,Me2002}. The problem of finding the most general $N$-soliton solution and their interactions has recently attracted a great deal of attention. In~\cite{BPPPr2001a} a sort of dressing method was applied to ``superimpose'' $N$-solitons to a generic, smooth and decaying background and to obtain the corresponding Jost solutions. There, the two soliton case was studied in details, showing that solitons can interact inelastically and that they can be created and annihilated. By using a finite dimensional version of the Sato theory for the KP hierarchy~\cite{Sato1981}, in a series of papers~\cite{BK,BC,BC2,B07,ChK1}, it was shown that the general $N$-soliton solution can be written in term of $\tau$-functions and their structure was studied in details, showing that they exhibit nontrivial spatial interaction patterns, resonances and web structures. A survey of these results is given in~\cite{ChK2} and applications to shallow water waves in~\cite{ChK2,BM09}. In~\cite{BPPP2009} solutions corresponding to $N$ solitons ``superimposed'' to a generic smooth decaying background and the corresponding Jost solutions were constructed by means of twisting transformations.

A spectral theory of the heat operator~(\ref{heatop}) that also includes multisolitons has to be built. In \cite{BPPP2002} the inverse scattering transform for a perturbed one-soliton potential was derived. In~\cite{Villaroel} the initial value problem for the KPII equation with data not decaying along a line was linearized. However, the case of $N$-solitons is still open. In solving the analogous problem for the nonstationary Schr\"{o}dinger operator, associated to the KPI equation, the extended resolvent approach was introduced~\cite{BPP2006b}. Accordingly, in order to solve the spectral problem for the heat operator, when the potential $u(x)$ describes $N$ solitons, one needs to find among different procedures used in deriving these potentials just that one that can be exploited in building the corresponding extended resolvent. Therefore, one needs to explore these different available procedures and their interrelation, which is the goal of this paper.

The paper is organized as follows. In section 2 we introduce basic notations and sketch some basic topics of the standard scattering problem of operator~(\ref{heatop}) for the case of decaying potential. In section 3 we review the multisoliton potentials constructed in~\cite{BPPPr2001a}, solving a $\overline{\partial}$-problem given by a rational transformation of generic spectral data. In section 4 we consider the multisoliton potentials obtained in~\cite{BPPP2009} by means of twisting transformations and we study their connection with the potentials given in section 3. In section 5 we derive some alternative, more symmetric representations of the pure solitonic potentials, and we show that they coincide with the representation obtained in~\cite{BK,BC,BC2,B07,ChK1} by using the $\tau$-function approach. In section 6 a symmetric representation for the Jost solutions is also obtained, showing that they can be obtained by a Miwa shift~\cite{Miwa}. Finally, we discuss the invariance properties of the multisoliton potentials with respect to transformations of the soliton parameters.

\section{Background theory}

The KPII can be expressed as compatibility condition of a Lax pair, $[\mathcal{L},\mathcal{T}]=0$, where $\mathcal{L}$ is the heat operator defined in~(\ref{heatop}) and $\mathcal{T}$ is given by
\begin{equation}
\mathcal{T}(x,\partial_{x},\partial_{t})=\partial_{t}+4
\partial_{x_{1}}^{3}-6u\partial_{x_{1}}-3u_{x_{1}}-
3\partial_{x_{1}}^{-1}u_{x_{2}}.  \label{Lax}
\end{equation}
Since the heat operator is not self-dual, one has to consider at the same time its dual $\mathcal{L}^{d}\left( x,\partial_{x}\right)=\partial_{x_{2}}+\partial_{x_{1}}^{2}-u(x)$ and, then, introduce the Jost solution $\Phi(x,\bk)$ and the dual Jost solution $\Psi(x,\bk)$ obeying equations
\begin{equation}
\mathcal{L}(x,\partial_{x})\Phi (x,\bk)=0,\qquad \mathcal{L}^{d}(x,\partial_{x})\Psi (x,\bk)=0,  \label{1.2}
\end{equation}
where $\bk$ is an arbitrary complex variable, playing the role of a spectral parameter.

It is convenient to normalize the Jost and dual Jost solutions multiplying them by an exponential function as follows
\begin{equation}
\chi (x,\bk)=e^{i\bk x_{1}+\bk^{2}x_{2}}\Phi (x,\bk),\qquad \xi (x,\bk)=e^{-i
\bk x_{1}-\bk^{2}x_{2}}\Psi (x,\bk).  \label{chiPhi}
\end{equation}
These functions satisfy the differential equations
\begin{subequations}
\label{chi-xi}
\begin{align}
& (-\partial_{x_{2}}+\partial_{x_{1}}^{2}-2i\bk\partial_{x_{1}}-u(x))\chi(x,
\bk)=0,  \label{eq-chi'} \\
& (\partial_{x_{2}}+\partial_{x_{1}}^{2}+2i\bk\partial_{x_{1}}-u(x))\xi(x,\bk
)=0,  \label{eq-xi'}
\end{align}
and are chosen to obey normalization conditions at $\bk$-infinity
\end{subequations}
\begin{equation}
\lim_{\bk\rightarrow \infty}\chi (x,\bk)=1,\qquad \lim_{\bk
\rightarrow\infty}\xi (x,\bk)=1,  \label{asymptk}
\end{equation}
so that, by~(\ref{chi-xi}), they are related to the potential $u(x)$ of the heat equation by means of the relations
\begin{equation}
u(x)=-2i\lim_{\bk\rightarrow \infty}\bk\partial_{x_{1}}\chi (x,\bk)=2i\lim_{
\bk\rightarrow \infty}\bk\partial_{x_{1}}\xi (x,\bk).  \label{asymptk1}
\end{equation}
The reality of the potential $u(x)$, that we always assume here, is equivalent to the conjugation properties
\begin{equation}
\overline{\chi (x,\bk)}=\chi (x,-\overline{\bk}),\qquad \overline{\xi (x,\bk)
}=\xi (x,-\overline{\bk}).  \label{chixiconj}
\end{equation}
In the case of a potential $u(x)\equiv u_{0}(x)$ rapidly decaying at spatial infinity, according to~\cite{BarYacoov,Lipovsky,Wickerhauser,Grinevich0}, the main tools in building the spectral theory of the operator~(\ref{heatop}), as in the one dimensional case, are the integral equations, whose solutions define the related normalized Jost solution $\chi_{0}(x,\bk)$, and $\xi_{0}(x,\bk)$, i.e.,
\begin{subequations}
\label{inteqs}
\begin{align}
\chi_{0}(x,\bk)& =1+\int dx^{\prime}\,G_{0}(x-x^{\prime},\bk
)u_{0}(x^{\prime})\chi_{0}(x^{\prime},\bk),  \label{4} \\
\xi_{0}(x,\bk)& =1+\int dx^{\prime}\,G_{0}(x^{\prime}-x,\bk
)u_{0}(x^{\prime})\xi_{0}(x^{\prime},\bk),  \label{5}
\end{align}
where
\end{subequations}
\begin{equation}
G_{0}(x,\bk)=-\frac{\sgn{x_{2}}}{2\pi}\int d\alpha \,\theta (\alpha(\alpha
-2\bk_{\Re})x_{2})e^{i\alpha x_{1}- \alpha (\alpha -2\bk)x_{2}},  \label{6}
\end{equation}
is the Green's function of the bare heat operator, $(-\partial_{x_{2}}+\partial_{x_{1}}^{2})G_{0}(x,\bk)=\delta (x)$.

Thanks to~(\ref{inteqs}), the functions $\chi_{0}$ and $\xi_{0}$ have the following asymptotic behaviour on the $x$-plane
\begin{equation}
\lim_{x\rightarrow \infty}\chi_{0}(x,\bk)=1,\qquad
\lim_{x\rightarrow\infty}\xi_{0}(x,\bk)=1.  \label{asymptx}
\end{equation}

However, if the potential $u(x)$ does not decay at spatial infinity, as it is the case when line soliton solutions are considered, the integral equations~(\ref{inteqs}) are ill-defined and one needs a more general approach. A spectral theory of the KPII equation that also includes solitons has been investigated using the resolvent approach. In this framework it was possible to develop the inverse scattering transform for a solution describing one soliton on a generic background~\cite{BPPP2002}, and to study the existence of the (extended) resolvent for (some) multisoliton solutions~\cite{BPPP2009}. The general theory is, to some extent, still a work in progress. In this paper, however, we consider only different approaches to the construction of soliton solutions and the corresponding Jost solutions.

\section{Multisoliton potentials via dressing method}

In~\cite{BPPPr2001a} we used a sort of dressing method to construct a potential $u(x)$ describing $N$ solitons superimposed to a generic background potential $u_{0}(x)$. More precisely, we considered a potential $u(x)$ with spectral data obtained by a rational transformation of the spectral data of a generic background $u_{0}(x)$ and we transformed the problem of finding this $u(x)$ into a $\overline{\partial}$-problem on the corresponding Jost solutions, which was solved by a dressing procedure. The rational transformation was chosen to depend on $N$ pairs of real distinct parameters $a_{j}$, $b_{j}$ ($j=1,\dots ,N)$ and the Jost solutions $\Phi (x,\bk)$ and $\Psi (x,\bk)$ were required to obey suitable analyticity properties, normalization like in~(\ref{chiPhi}),~(\ref{asymptk}) and conjugation property~(\ref{chixiconj}).

The transformed potential $u(x)$ is given by the following formula
\begin{equation}
u(x)=u_{0}(x)-2\partial_{x_{1}}^{2}\det [C+\mathcal{F}(x)],  \label{square1}
\end{equation}
where $C$ is a real $N\times N$ constant matrix and $\mathcal{F}(x)$ is an $N\times N$ matrix function of $x$ with entries
\begin{equation}
\mathcal{F}_{lj}(x)=\int\limits_{\substack{ (b_{l}-a_{j})\infty  \\
y_{2}=x_{2}}}^{x_{1}}dy_{1}\,\Psi_{0}(y,ib_{l})\Phi_{0}(y,ia_{j}),\quad
l,j=1,\dots ,N,  \label{F1}
\end{equation}
where $\Phi_{0}(x,\bk)$ and $\Psi_{0}(x,\bk)$ are the Jost and dual Jost solutions of equations~(\ref{1.2}) with potential $u_{0}(x)$. Notice that the matrix elements of $\Fo(x)$ are given in terms of values of the so-called Cauchy--Baker--Akhiezer function~\cite{Grinevich} at points $a_{j}$ and $b_{l}$ with $j,l=1,\ldots ,N$. In~\cite{BPPPr2001a} we have shown that the matrix $C$ does not need to be either regular, or diagonal. We also mentioned that in order to obtain a real potential it is enough to consider a more general situation, with complex parameters $a_{j}$ and $b_{j}$ such that
\begin{equation}
a_{j}=\bar{a}_{\pi_{a}(j)},\qquad b_{j}=\bar{b}_{\pi_{b}(j)},\quad j=1,\dots
,N,  \label{cc}
\end{equation}
where the bar denotes the complex conjugate and $\pi_{a}$, $\pi_{b}$ are some permutations of the indices (with a proper modification of the constraints on the constant matrix $C$). However, this case is essentially more complicated to be investigated and we do not consider it here.

If, a posteriori, the background $u_{0}$ is taken to be identically zero, then the eigenfunctions in~(\ref{F1}) are pure exponential functions, i.e.,
\begin{equation*}
\Phi_{0}(x,ia_{j})=e^{A_{j}(x)},\qquad \Psi_{0}(x,ib_{l})=e^{-B_{l}(x)},
\end{equation*}
where for the future convenience we introduced
\begin{equation}
A_{j}(x)=a_{j}x_{1}+a_{j}^{2}x_{2},\qquad
B_{l}(x)=b_{l}x_{1}+b_{l}^{2}x_{2},\quad j,l=1,2,\dots ,N.  \label{AjBjx0}
\end{equation}
Then~(\ref{F1}) takes the form
\begin{equation}
\mathcal{F}_{lj}(x)=e^{-B_{l}(x)}\Lambda_{lj}e^{A_{j}(x)},  \label{Flj}
\end{equation}
with $\Lambda$ the Cauchy matrix
\begin{equation}
\Lambda_{lj}=\frac{1}{a_{j}-b_{l}}.  \label{Cauchymatr}
\end{equation}

Notice that in~\cite{BPPPr2001a} the time dependence was not specified, but this simply amounts to taking into account the time dependence of the original Jost solutions $\Phi_{0}(x,\bk)$ and $\Psi_{0}(x,\bk)$ fixed by the choice of the second Lax operator~(\ref{Lax}). In the case of $u_{0}(x,t)\equiv 0$ this means that we have to use
\begin{equation}
A_{j}(x,t)=a_{j}x_{1}+a_{j}^{2}x_{2}-4a_{j}^{3}t,\qquad
B_{l}(x,t)=b_{l}x_{1}+b_{l}^{2}x_{2}-4b_{l}^{3}t,  \label{AjBjxt}
\end{equation}
instead of~(\ref{AjBjx0}).

In~\cite{BPPPr2001a} we addressed the problem of regularity of the potentials, as well as of their asymptotic behaviour, only for the case $N=2$, i.e., for $2$-soliton potentials. In this particular case, we were able to formulate the conditions on the $2\times 2$ matrix $C$ that guarantee the regularity of the potential $u(x)$, and we showed that at large distances in the $x$-plane the potential decays exponentially fast except along certain specific rays, where it has a solitonic one dimensional behaviour of the form~(\ref{1-sol}). The classification of the $2$-soliton potentials obtained in~\cite{BPPPr2001a} was then successively obtained and generalized to the case of $N$-soliton in~\cite{ChK1,ChK2}.

\section{Multisoliton potentials via twisting transformations}

\subsection{Equivalence of the $N$-soliton potentials derived in~
\protect\cite{BPPPr2001a} and in \protect\cite{BPPP2009}}

In this section we consider pure soliton potentials of the heat equation obtained by twisting transformations. The details of the construction of the twisting operators and of the corresponding potentials can be found in~\cite{BPPP2009}. The transformation of a generic (smooth, decaying at infinity) background potential $u_{0}(x)$ into a new potential $u(x)$, which describes $N$-solitons ``superimposed'' to the background $u_{0}(x)$, is parameterized by two sets of real parameters $\{a_{1},\ldots ,a_{N_{a}}\}$ and $\{b_{1},\ldots ,b_{N_{b}}\}$, which we assume all distinct, and an $N_{a}\times N_{b}$ real matrix $c$. In this case we allow $N_{a}$ and $N_{b}$ to be not necessarily equal, but
\begin{equation}
N_{a},N_{b}\geq 1,  \label{Nab}
\end{equation}
and denote $N=\max \{N_{a},N_{b}\}$. The pure $N$-soliton potential follows directly from the expressions derived in~\cite{BPPP2009} by taking $u_{0}(x)\equiv 0$. Precisely, we have
\begin{equation}
u(x)=-2\partial _{x_{1}}^{2}\log \tau _{1}(x),  \label{starting}
\end{equation}
with
\begin{equation}
\tau _{1}(x)=\det (E_{N_{a}}+c\Fo(x))\equiv \det (E_{N_{b}}+\Fo(x)c),
\label{ex3}
\end{equation}
where we introduced the $N_{a}\times N_{a}$ (resp.\ $N_{b}\times N_{b}$) identity matrix $E_{N_{a}}$ (resp.\ $E_{N_{b}}$), the diagonal matrices
\begin{equation}
e^{A(x)}=\diag\{e^{A_{j}(x)}\}_{j=1}^{N_{a}},\qquad e^{-B(x)}=\diag
\{e^{-B_{l}(x)}\}_{l=1}^{N_{b}},  \label{eAB}
\end{equation}
see~(\ref{AjBjx0}), and the $N_{b}\times N_{a}$ matrix function $\Fo(x)=\Vert \Fo_{lj}(x)\Vert _{l=1,\ldots ,N_{b}}^{j=1,\ldots ,N_{a}}$, where notation~(\ref{Flj}) was used, so that
\begin{equation}
\Fo(x)=e^{-B(x)}\Lambda e^{A(x)},  \label{FBA}
\end{equation}
with $\Lambda $ a $N_{b}\times N_{a}$ constant matrix with elements given
in~(\ref{Cauchymatr}).

Let us show now that the potentials given in~(\ref{square1}) in terms of $N\times {N}$ matrices $C$ and $\mathcal{F}$ coincide with those obtained by the twisting transformations and expressed by means of matrices $c$ and $\mathcal{F}$ in~(\ref{ex3}). This is obvious when $N_{a}=N_{b}$ and matrices $C$ and $c$ are nonsingular. Then, it is enough to put $c=C^{-1} $ so that the determinants in~(\ref{square1}) and~(\ref{ex3}) are equal up to an unessential constant factor.

The generic situation is a bit more complicated. In order to make explicit the size of the matrices involved we use here notation $A_{N}$ for a $N\times {N}$ matrix $A$. Let us consider, for definiteness, the case $N_{a}\leq {N_{b}}\equiv N$. Let $c_{N}$ denote the $N\times {N}$ matrix constructed by adding $N_{b}-N_{a}$ zero rows to the matrix $c$. Taking into account that the product $\mathcal{F}c$ in the second equality in~(\ref{ex3}) is an $N\times {N}$-matrix, it is clear that $(\mathcal{F)}_{N}c_{N}=\mathcal{F}c$, where
\begin{equation}
\mathcal{F}_{N}=(e^{-B})_{N}\Lambda_{N}(e^{A})_{N}  \label{FN}
\end{equation}
and where the parameters $a_{j}$ with $j=N_{a}+1,N_{a}+2,\ldots ,N$ in $(e^{A})_{N}$ and in $\Lambda_{N}$ can be chosen arbitrarily, provided they are different from the parameters $b_{l}$. It is well known that the Cauchy matrix $\Lambda_{N} $ is invertible and that $\Lambda_{N}^{-1}=G_{N}\widetilde{\Lambda}_{N}F_{N}$, where the matrix $\widetilde{\Lambda}_{N}$ is obtained from the matrix $\Lambda_{N}$ by renaming $a_{j}$ the $b_{j}$ and vice-versa and where $G_{N}$ and $F_{N}$ are two invertible diagonal matrices. Thus, for the second determinant in~(\ref{ex3}) we have
\begin{align*}
&\det (E_{N_{b}}+\mathcal{F}c)=\det (E_{N}+\Fo_{N}c_{N})\\
&\qquad=\det(e^{-B}\Lambda e^{A}GF)_{N} \det \bigl((e^{-A})_{N}\widetilde{\Lambda}
_{N}(e^{B})_{N}+G_{N}^{-1}c_{N}^{}F_{N}^{-1}\bigr).
\end{align*}
The factor $\det (e^{-B}\Lambda e^{A}GF)_{N}$ does not contribute to the potential, due to the derivative in~(\ref{starting}). Moreover, thanks to~(\ref{Flj}) and~(\ref{FN}), the $N\times N$ matrix $e^{-A(x)}\widetilde{\Lambda}e^{B(x)}$ coincides with the $N\times {N}$-matrix $\Fo(x)$ in ~(\ref{Flj}), once parameters $a_{j}$ and $b_{j}$ are exchanged. Therefore, relations~(\ref{ex3}) and~(\ref{starting}) indeed generate the same potential provided
\begin{equation}
c_{N}=G_{N}CF_{N}  \label{x6}
\end{equation}
and parameters $a_{j}$ are renamed $b_{j}$ and vice-versa.

\begin{remark}
In the case of a nonzero background potential $u_{0}(x)$ one can prove that the two potentials obtained by the two approaches are equivalent, up to an additional term $\partial_{x_{1}}^{2}\log \det \Fo_{N}(x)$.
\end{remark}

\begin{remark}
Here and below we omitted the time dependence as it can be easily switched on by using~(\ref{AjBjxt}) instead of~(\ref{AjBjx0}).
\end{remark}

\subsection{Regularity conditions for the potential}

In view of the discussion on the necessary and sufficient conditions required to guarantee the regularity of the multisoliton solutions of the equations in the hierarchy of the KPII equation, we recover here, in an equivalent formulation, sufficient conditions for regularity of multisoliton potentials of the heat equation and, then, of multisoliton  solutions of KPII, already given in~\cite{BC,ChK2}.

Let us consider again the second equality in~(\ref{ex3}). We have
\begin{equation}
\det (E_{N_{b}}+{\Fo}c)=\sum_{n=0}^{N_{b}}\;\sum_{1\leq
l_{1}<l_{2}<\dots<l_{n}\leq N_{b}}({\Fo}c)\!\left( \!\!
\begin{array}{c}
l_{1},l_{2},\dots ,l_{n} \\
l_{1},l_{2},\dots ,l_{n}
\end{array}
\!\!\!\right) .  \label{detDelta}
\end{equation}
Here and in the following $A\left( \!\!
\begin{array}{c}
l_{1},l_{2},\dots ,l_{n} \\
l_{1},l_{2},\dots ,l_{n}
\end{array}
\!\!\!\right)$ denotes the minor of the matrix $A$ obtained by selecting rows $l_1,l_1,\dots,l_n$ and columns $j_1,j_2,\dots,j_n$. The principal minor of the product ${\Fo}c$ in~(\ref{detDelta}) can be written by the Binet--Cauchy formula as
\begin{align}
({\Fo}c)\!\left( \!\!
\begin{array}{c}
l_{1},l_{2},\dots ,l_{n} \\
l_{1},l_{2},\dots ,l_{n}
\end{array}
\!\!\!\right) & =\sum_{1\leq j_{1}<j_{2}<\dots <j_{n}\leq N_{a}}
\left(\prod_{m=1}^{n}e^{-B_{l_{m}}(x)}\right)
\left(\prod_{m=1}^{n}e^{A_{j_{m}}(x)}\right)  \notag \\
& \times \Lambda \!\left( \!\!
\begin{array}{c}
l_{1},l_{2},\dots ,l_{n} \\
j_{1},j_{2},\dots ,j_{n}
\end{array}
\!\!\!\right) c\!\left( \!\!
\begin{array}{c}
j_{1},j_{2},\dots ,j_{n} \\
l_{1},l_{2},\dots ,l_{n}
\end{array}
\!\!\!\right) .  \label{expos}
\end{align}
Taking into account that $\Fo$ is a $N_{b}\times {N_{a}}$-matrix and $c$ a $N_{a}\times {N_{b}}$-matrix we get that all minors of ${\Fo}c$ with $n>\min{N_{a},N_{b}}$ are zero.

Recalling that the term corresponding to $n=0$ in~(\ref{detDelta}) is $1$, i.e., greater than zero, we deduce that a sufficient condition for having a regular solution is that the real matrix $c$ satisfies the following characterization conditions
\begin{equation}
\Lambda \!\left( \!\!
\begin{array}{c}
l_{1},l_{2},\dots ,l_{n} \\
j_{1},j_{2},\dots ,j_{n}
\end{array}
\!\!\!\right) c \!\left( \!\!
\begin{array}{c}
j_{1},j_{2},\dots ,j_{n} \\
l_{1},l_{2},\dots ,l_{n}
\end{array}
\!\!\!\right) \geq 0,  \label{condition}
\end{equation}
for any $1\leq n\leq \min \{N_{a},N_{b}\}$ and all minors, i.e., any choice of $1\leq j_{1}<j_{2}<\dots <j_{n}\leq N_{a}$ and $1\leq l_{1}<l_{2}<\dots l_{n}\leq N_{b}$. In order to get a nontrivial potential under substitution of $\tau_{1}$ in~(\ref{starting}) it is necessary then to impose the condition that at least one of inequalities in~(\ref{condition}) is strict.

Now, since any submatrix of a Cauchy matrix $\Lambda $ is itself a Cauchy matrix, for an arbitrary minor of $\Lambda $ we have
\begin{equation}
\Lambda \!\left( \!\!
\begin{array}{c}
l_{1},l_{2},\dots ,l_{p} \\
j_{1},j_{2},\dots ,j_{p}
\end{array}
\!\!\!\right) =\frac{\displaystyle\prod_{1\leq i<m\leq
p}(a_{j_{i}}-a_{j_{m}})\left( b_{l_{m}}-b_{l_{i}}\right)} {\displaystyle
\prod_{i,m=1}^{p}(a_{j_{i}}-b_{j_{m}})}.  \label{minors_Cauchy}
\end{equation}
Then, we deduce that there are orders of the parameters $a$'s and $b$'s for which all minors of the matrix $\Lambda $ are positive. For instance, if
\begin{equation}
a_{1}<a_{2}<\dots <a_{N_{a}-1}<a_{N_{a}}<b_{1}<b_{2}<\dots
<b_{N_{b}-1}<b_{N_{b}},  \label{bachoice}
\end{equation}
the regularity conditions~(\ref{condition}) become
\begin{equation}
c\!\left( \!\!
\begin{array}{c}
l_{1},l_{2},\dots ,l_{n} \\
j_{1},j_{2},\dots ,j_{n}
\end{array}
\!\!\!\right) \geq 0,
\end{equation}
i.e., all minors of the matrix $c$ must be non negative. These matrices are called totally nonnegative matrices \cite{G2002}. Obviously one obtains the same result by using the first equality in~(\ref{ex3}) and the Binet--Cauchy formula enables to check directly that the two determinants in~(\ref{ex3}) are equal.

If we consider the multi-time multisoliton solution of the entire hierarchy related to the KPII equation obtained by the Miwa shift (see \cite{Miwa}), as reported in~(\ref{Kn}), we get again an expansion of the form~(\ref{expos}), where the exponents are now independent and we conclude that the regularity condition~(\ref{condition}) is also necessary.

\section{Equivalence with the $\protect\tau $-function representation}

\subsection{Determinant representations for the potential and Jost solutions}

In~\cite{BPPP2009} we derived the transformed potential $u(x)$ (see~(\ref{starting} and~(\ref{ex3})) and the Jost solutions of the direct and dual problems~(\ref{1.2}) related to this potential. The corresponding normalized Jost solutions are expressed as
\begin{subequations}
\label{PhiPsi4}
\begin{align}
\chi (x,\bk)& =1+\sum_{j=1}^{N_{a}}\sum_{l,l^{\prime
}=1}^{N_{b}}e^{A_{j}(x)}c_{jl^{\prime }}(E_{N_{b}}+\Fo(x)c)_{l^{\prime
}l}^{-1}\frac{e^{-B_{l}(x)}}{b_{l}+i\bk}  \notag \\
& \equiv 1+\sum_{j,j^{\prime
}=1}^{N_{a}}\sum_{l=1}^{N_{b}}e^{A_{j}(x)}(E_{N_{a}}+c\Fo(x))_{jj^{\prime
}}^{-1}\frac{c_{j^{\prime }l}e^{-B_{l}(x)}}{b_{l}+i\bk},  \label{Phi4} \\
\xi (x,\bk)& =1-i\sum_{j=1}^{N_{a}}\sum_{l,l^{\prime }=1}^{N_{b}}\frac{
e^{A_{j}(x)}}{a_{j}+i\bk}c_{jl}(E_{N_{b}}+\Fo(x)c)_{ll^{\prime
}}^{-1}e^{-B_{l}(x)}  \notag \\
& \equiv 1-\sum_{j,j^{\prime }=1}^{N_{a}}\sum_{l=1}^{N_{b}}\frac{
e^{A_{j}(x)}}{a_{j}+i\bk}(E_{N_{a}}+c\Fo(x))_{jj^{\prime }}^{-1}c_{j^{\prime
}l}e^{-B_{l}(x)}.  \label{Psi4}
\end{align}
\end{subequations}
In analogy to~(\ref{ex3}), all relations are given in two equivalent forms. While it is enough to keep only one of them, we continue to consider both of them to highlight the symmetry property of the whole construction with respect to the numbers $N_{a}$ and $N_{b}$, which play the role of topological charges.

In order to obtain a representation of ~(\ref{ex3}) and~(\ref{PhiPsi4}) in terms of $\tau $-functions we need to perform some simple algebraic operations. First, thanks to the standard identity for the determinant of a bordered matrix, we can rewrite~(\ref{PhiPsi4}) as
\begin{equation}
\chi (x,\bk)=\frac{\tau _{1,\chi }(x,\bk)}{\tau _{1}(x)},\qquad \xi (x,\bk)=
\frac{\tau _{1,\xi }(x,\bk)}{\tau _{1}(x)},  \label{sym4.10}
\end{equation}
where $\tau _{1}$ is given in~(\ref{ex3}) and
\begin{subequations}
\label{sym56}
\begin{align}
\tau _{1,\chi }(x,\bk)& =\det \left(
\begin{array}{cc}
E_{N_{b}}+\Fo(x)c & \displaystyle\frac{-e^{-B_{\ast }(x)}}{b_{\ast }+i\bk} \\
\displaystyle\sum_{j=1}^{N_{a}}e^{A_{j}(x)}c_{j\ast } & 1
\end{array}
\right)  \label{sym5:2} \\
& \equiv \det \left(
\begin{array}{cc}
E_{N_{a}}+c\Fo(x) & -\displaystyle\sum_{l=1}^{N_{b}}\frac{c_{\ast l}\
e^{-B_{l}(x)}}{b_{l}+i\bk} \\
e^{A_{\ast }(x)} & 1
\end{array}
\right) ,  \label{sym5:1} \\
\tau _{1,\xi }(x,\bk)& =\det \left(
\begin{array}{cc}
E_{N_{b}}+\Fo(x)c & e^{-B_{\ast }(x)} \\
\displaystyle\sum_{j=1}^{N_{a}}\frac{e^{A_{j}(x)}}{a_{j}+i\bk}c_{j\ast } & 1
\end{array}
\right)   \label{sym6:2} \\
& \equiv \det \left(
\begin{array}{cc}
E_{N_{a}}+c\Fo(x) & \displaystyle\sum_{l=1}^{N_{b}}c_{\ast l}\ e^{-B_{l}(x)}
\\
\displaystyle\frac{e^{A_{\ast }(x)}}{a_{\ast }+i\bk} & 1
\end{array}
\right) .  \label{sym6:1}
\end{align}
\end{subequations}
In the above formulas, in the row and column bordering matrices $E_{N_{b}}+\Fo c$ and $E_{N_{a}}+c\Fo$ we use the subscript ${\ast }$ to denote an index running, respectively, from $1$ to $N_{b}$ and from $1$ to $N_{a}$. Let us notice that the function $\tau _{1}$ in \eqref{ex3} can also be obtained as a limiting value, i.e.,
\begin{equation}
\tau _{1}(x)=\lim_{\bk\rightarrow \infty }\tau _{1,\chi }(x,\bk)=\lim_{\bk
\rightarrow \infty }\tau _{1,\xi }(x,\bk).  \label{sym7}
\end{equation}
Using~(\ref{chiPhi}) to go back from~(\ref{sym4.10}) and~(\ref{sym56}) to the Jost solutions $\Phi (x,\bk)$ and $\Psi (x,\bk)$, one can check that they have poles, respectively, at $\bk=ib_{l}$ ($l=1,\ldots ,N_{b}$) and $\bk=ia_{j}$ ($j=1,\ldots ,N_{a}$) with residua
\begin{equation}
\Phi_{b_{l}}(x)=\res_{\bk=ib_{l}}\Phi (x,\bk),\qquad \Psi_{a_{j}}(x)=\res_{
\bk=ia_{j}}\Psi (x,\bk),  \label{res}
\end{equation}
which obey the relations
\begin{equation}
\Phi_{b_{l}}(x)=-i\sum_{j=1}^{N_{a}}\Phi
(x,ia_{j})c_{jl},\qquad\Psi_{a_{j}}(x)=i\sum_{l=1}^{N_{b}}c_{jl}\Psi
(x,ib_{l}).  \label{PhiPsi}
\end{equation}
Notice that these equations, together with the requirement of analyticity and the normalization condition~(\ref{asymptx}), allow one to reconstruct the normalized Jost solutions in~(\ref{sym4.10}) and~(\ref{sym56}) and, then, via~(\ref{asymptk1}) the potential $u(x)$ in~(\ref{starting}) and~(\ref{ex3}).

Next, we rewrite~(\ref{sym56}) as
\begin{align*}
\tau _{1,\chi }(x,\bk)& =\left( \prod_{l=1}^{N_{b}}\frac{e^{-B_{l}(x)}}{
b_{l}+i\bk}\right) \det \left(
\begin{array}{cc}
(b_{\ast }+i\bk)[e^{B(x)}+\Lambda e^{A(x)}c] & -1_{\ast } \\
\displaystyle\sum_{j=1}^{N_{a}}e^{A_{j}(x)}c_{j\ast } & 1
\end{array}
\right) \\
& \equiv \left( \prod_{j=1}^{N_{a}}e^{A_{j}(x)}\right) \det \left(
\begin{array}{cc}
e^{-A(x)}+ce^{-B(x)}\Lambda & -\displaystyle\sum_{l=1}^{N_{b}}\frac{c_{\ast
l}\ e^{-B_{l}(x)}}{b_{l}+i\bk} \\
1_{\ast } & 1
\end{array}
\right) , \\
\tau _{1,\xi }(x,\bk)& =\left( \prod_{l=1}^{N_{b}}e^{-B_{l}(x)}\right) \det
\left(
\begin{array}{cc}
e^{B(x)}+\Lambda e^{A(x)}c & 1_{\ast } \\
\displaystyle\sum_{j=1}^{N_{a}}\frac{e^{A_{j}(x)}}{a_{j}+i\bk}c_{j\ast } & 1
\end{array}
\right) \\
& \equiv \left( \prod_{j=1}^{N_{a}}\frac{e^{A_{j}(x)}}{a_{j}+i\bk}\right)
\det \left(
\begin{array}{cc}
\lbrack e^{-A(x)}+ce^{-B(x)}\Lambda ](a_{\ast }+i\bk) & \displaystyle
\sum_{l=1}^{N_{b}}c_{\ast l}e^{-B_{l}(x)} \\
1_{\ast } & 1
\end{array}
\right) .
\end{align*}
By elementary transformations of the matrices on the right-hand sides, we can reduce them to a form where rows and columns with all elements equal to $1$ or $-1$ are transformed into rows and columns with all elements equal to $0$ except $1$ at the last place.

In order to present the result of these transformations we introduce
\begin{subequations}
\label{sol67}
\begin{align}
\tau _{2,\chi }(x,\bk)& =\left( \prod_{l=1}^{N_{b}}e^{B_{l}(x)}(b_{l}+i\bk
)\right) \tau _{1,\chi }(x,\bk),  \label{sol6:2} \\
\tau _{2,\chi }^{\prime }(x,\bk)& =\left( \prod_{j=1}^{N_{a}}\frac{
e^{-A_{j}(x)}}{a_{j}+i\bk}\right) \tau _{1,\chi }(x,\bk),  \label{sol6:1} \\
\tau _{2,\xi }(x,\bk)& =\left( \prod_{l=1}^{N_{b}}\frac{e^{B_{l}(x)}}{
b_{l}+i\bk}\right) \tau _{1,\xi }(x,\bk),  \label{sol7:2} \\
\tau _{2,\xi }^{\prime }(x,\bk)& =\left(
\prod_{l=1}^{N_{a}}e^{-A_{j}(x)}(a_{j}+i\bk)\right) \tau _{1,\xi }(x,\bk).
\label{sol7:1}
\end{align}
\end{subequations}
Then,
\begin{subequations}
\label{sol89}
\begin{align}
\tau _{2,\chi }(x,\bk)& =\det \left( \delta _{ll^{\prime }}(b_{l}+i\bk
)e^{B_{l}(x)}+\sum_{j=1}^{N_{a}}\frac{a_{j}+i\bk}{a_{j}-b_{l}}
e^{A_{j}(x)}c_{jl^{\prime }}\right) _{l,l^{\prime }=1}^{N_{b}},  \label{sol8:2} \\
\tau _{2,\chi }^{\prime }(x,\bk)& =\det \left( \delta _{jj^{\prime }}\frac{
e^{-A_{j}(x)}}{a_{j}+i\bk}+\sum_{l=1}^{N_{b}}\frac{c_{jl}e^{-B_{l}(x)}}{
(a_{j^{\prime }}-b_{l})(b_{l}+i\bk)}\right)_{j,j^{\prime
}=1}^{N_{a}} ,  \label{sol8:1} \\
\tau _{2,\xi }(x,\bk)& =\det \left( \delta _{ll^{\prime }}\frac{e^{B_{l}(x)}
}{b_{l}+i\bk}+\sum_{j=1}^{N_{a}}\frac{e^{A_{j}(x)}c_{jl^{\prime }}}{
(a_{j}-b_{l})(a_{j}+i\bk)}\right) _{l,l^{\prime }=1}^{N_{b}},  \label{sol9:2}
\\
\tau _{2,\xi }^{\prime }(x,\bk)& =\det \left( \delta _{jj^{\prime }}(a_{j}+i
\bk)e^{-A_{j}(x)}+\sum_{l=1}^{N_{b}}\frac{c_{jl}(b_{l}+i\bk)}{a_{j^{\prime
}}-b_{l}}e^{-B_{l}(x)}\right) _{j,j^{\prime }=1}^{N_{a}}.  \label{sol9:1}
\end{align}
\end{subequations}
Let us now introduce the limits
\begin{align*}
\tau _{2}(x)& =\lim_{\bk\rightarrow \infty }(i\bk)^{-N_{b}}\tau _{2,\chi }(x,
\bk)=\lim_{\bk\rightarrow \infty }(i\bk)^{N_{b}}\tau _{2,\xi }(x,\bk), \\
\tau _{2}^{\prime }(x)& =\lim_{\bk\rightarrow \infty }(i\bk)^{N_{a}}\tau
_{2,\chi }^{\prime }(x,\bk)=\lim_{\bk\rightarrow \infty }(i\bk)^{-N_{a}}\tau
_{2,\xi }^{\prime }(x,\bk),
\end{align*}
that, thanks to~(\ref{sol89}), have the following explicit expressions
\begin{subequations}
\label{sol12a}
\begin{align}
\tau _{2}(x)& =\det \left( \delta _{ll^{\prime
}}e^{B_{l}(x)}+\sum_{j=1}^{N_{a}}\frac{e^{A_{j}(x)}c_{jl^{\prime }}}{
a_{j}-b_{l}}\right) _{l,l^{\prime }=1}^{N_{b}},  \label{sol12:2} \\
\tau _{2}^{\prime }(x)& =\det \left( \delta _{jj^{\prime
}}e^{-A_{j}(x)}+\sum_{l=1}^{N_{b}}\frac{c_{jl}e^{-B_{l}(x)}}{a_{j^{\prime
}}-b_{l}}\right) _{j,j^{\prime }=1}^{N_{a}}.  \label{sol12:1}
\end{align}
\end{subequations}
By~(\ref{sym7}) we have
\begin{equation}
\tau _{2}(x)=\left( \prod_{l=1}^{N_{b}}e^{B_{l}(x)}\right) \tau
_{1}(x),\qquad \tau _{2}^{\prime }(x)=\left(
\prod_{j=1}^{N_{a}}e^{-A_{j}(x)}\right) \tau _{1}(x).  \label{sol10}
\end{equation}
Both functions in~(\ref{sol12a}) are equivalent, in the sense that they generate the same potential
\begin{equation}
u(x)=-2\partial _{x_{1}}^{2}\log \tau _{2}(x)=-2\partial _{x_{1}}^{2}\log
\tau _{2}^{\prime }(x).  \label{solex3:1}
\end{equation}

For the functions $\chi (x,\bk)$ and $\xi (x,\bk)$ (see~(\ref{chiPhi})) we get by~(\ref{sym4.10}) and~(\ref{sol67})
\begin{subequations}
\label{sym34}
\begin{align}
\chi (x,\bk)& =\left( \prod_{l=1}^{N_{b}}(b_{l}+i\bk)^{-1}\right) \frac{
\tau _{2,\chi }(x,\bk)}{\tau _{2}(x)}\equiv \left(
\prod_{j=1}^{N_{a}}(a_{j}+i\bk)\right) \frac{\tau _{2,\chi }^{\prime }(x,\bk
)}{\tau _{2}^{\prime }(x)},  \label{sym3:1} \\
\xi (x,\bk)& =\left( \prod_{l=1}^{N_{b}}(b_{l}+i\bk)\right) \frac{\tau
_{2,\xi }(x,\bk)}{\tau _{2}(x)}\equiv \left( \prod_{j=1}^{N_{a}}(a_{j}+i\bk
)^{-1}\right) \frac{\tau _{2,\xi }^{\prime }(x,\bk)}{\tau _{2}^{\prime }(x)}
.  \label{sym4:1}
\end{align}
\end{subequations}
Notice that functions in~(\ref{sol89}) can be obtained from functions in~(\ref{sol12a}) by performing the following substitutions
\begin{subequations}
\label{sol11}
\begin{align}
\tau _{2}(x)& \rightarrow \tau _{2,\chi }(x,\bk)\quad \text{replacing}\quad
e^{A_{j}}\rightarrow e^{A_{j}}(a_{j}+i\bk),\quad e^{B_{l}}\rightarrow
e^{B_{l}}(b_{l}+i\bk)  \label{sol11:2} \\
\tau _{2}^{\prime }(x)& \rightarrow \tau _{2,\chi }^{\prime }(x,\bk)\quad
\text{replacing}\quad e^{A_{j}}\rightarrow e^{A_{j}}(a_{j}+i\bk),\quad
e^{B_{l}}\rightarrow e^{B_{l}}(b_{l}+i\bk),,  \label{sol11:1} \\
\tau _{2}(x)& \rightarrow \tau _{2,\xi }(x,\bk)\quad \text{replacing}\quad
e^{A_{j}}\rightarrow \frac{e^{A_{j}}}{a_{j}+i\bk},\quad
e^{B_{l}}\rightarrow \frac{e^{B_{l}}}{b_{l}+i\bk},  \label{sol11:4} \\
\tau _{2}^{\prime }(x)& \rightarrow \tau _{2,\xi }^{\prime }(x,\bk)\quad
\text{replacing}\quad e^{A_{j}}\rightarrow \frac{e^{A_{j}}}{a_{j}+i\bk}
,\quad e^{B_{l}}\rightarrow \frac{e^{B_{l}}}{b_{l}+i\bk}.  \label{sol11:3}
\end{align}
\end{subequations}
These rules enable to significantly shorten the list of formulas given below (see also Remark~\ref{Miwa}).

\begin{remark}
Notice that under the transformation $x\rightarrow -x$, $N_{a}\longleftrightarrow N_{b}$ and $a\longleftrightarrow b$ the function $\chi$ is transformed into $\xi $ and viceversa.
\end{remark}

\subsection{Symmetric representations for the potential and the comparison
with the $\protect\tau$-function approach}

Here we prove that the expression~(\ref{solex3:1}) for the potential is equivalent to that one obtained by means of the $\tau $-function approach in the series of papers quoted in the introduction and surveyed in \cite{ChK2}.

We already mentioned that the double representations for the potential and the Jost solutions derived above highlight the symmetric role played by the parameters $a_{j}$ and $b_{l}$. To better exploit this fact we introduce $N_{a}+N_{b}$ (real) parameters
\begin{equation}
\{\kappa_{1},\ldots
,\kappa_{N_{a}+N_{b}}\}=\{a_{1},\ldots,a_{N_{a}},b_{1}\ldots ,b_{N_{b}}\},
\label{sol12}
\end{equation}
and, by analogy with~(\ref{AjBjx0}), we introduce
\begin{equation}
K_{n}^{}(x)=\kappa_{n}^{}x_{1}^{}+\kappa_{n}^{2}x_{2}^{},\quad n=1,\ldots
,N_{a}^{}+N_{b}^{},  \label{An}
\end{equation}
so that $K_{n}(x)=A_{n}(x)$, $n=1,\ldots ,N_{a}$, and $
K_{n}(x)=B_{n-N_{a}}(x)$, $n=N_{a}+1,\ldots ,N_{a}+N_{b}$.

According to~(\ref{AjBjxt}) the time dependence is taken into account simply by adding a term $-4\kappa_{n}^{3}t$ to the rhs of~(\ref{An}).

Let $d$ denote an $N_{a}\times {N_{b}}$ real matrix with elements given in terms of the elements of the matrix $c$ as
\begin{equation}
d_{jl}=\prod_{l^{\prime }=1}^{N_{b}}(a_{j}-b_{l^{\prime
}})^{-1}c_{jl}\prod_{l^{\prime }=1,\,l^{\prime }\neq
l}^{N_{b}}(b_{l}-b_{l^{\prime }}),\quad j=1,\ldots ,N_{a},\quad l=1,\ldots
,N_{b}.  \label{sol13:2}
\end{equation}
Let us also introduce the constant $(N_{a}+N_{b})\times N_{b}$- and $N_{a}\times (N_{a}+N_{b})$-matrices $\Do$ and $\Do^{\,\prime}$ with, respectively, the following block structures
\begin{equation}
\Do=\left(
\begin{array}{c}
d \\
E_{N_{b}}
\end{array}
\right) ,\qquad \Do^{\,\prime}=\left( E_{N_{a}},-d\right) ,  \label{block}
\end{equation}
and the constant, diagonal, real $(N_{a}+N_{b})\times (N_{a}+N_{b})$-matrix
\begin{equation}
\gamma =\diag\left\{ \prod_{n^{\prime }=1,n^{\prime }\neq
n}^{N_{a}+N_{b}}(\kappa _{n}-\kappa _{n^{\prime }})^{-1},\quad n=1,\ldots
,N_{a}+N_{b}\right\} .  \label{d11:1}
\end{equation}

Let us prove now that with one more rescaling of $\tau _{2}(x)$ and $\tau_{2}^{\prime }(x)$:
\begin{subequations}
\label{sol15}
\begin{align}
& \tau (x)=\left( \prod_{1\leq l<l^{\prime}\leq N_{b}}(b_{l^{\prime }}-b_{l})\right)
\tau _{2}(x),  \label{sol15:2} \\
& \tau ^{\prime }(x)=\left( \prod_{1\leq {j}<j^{\prime }\leq {N_{a}}
}(a_{j}-a_{j^{\prime }})^{-1}\right) \left(
\prod_{j=1}^{N_{a}}\prod_{l=1}^{N_{b}}(a_{j}-b_{l})^{-1}\right) \tau
_{2}^{\prime }(x),  \label{sol15:1}
\end{align}
\end{subequations}
we get instead of~(\ref{sol12a})
\begin{equation}
\tau (x)=\frac{\det \bigl(\Ko e^{K(x)}\Do\bigr)}{\displaystyle\prod_{1\leq
l<l^{\prime }\leq N_{b}}(b_{l}-b_{l^{\prime }})},\qquad \tau ^{\prime }(x)=
\frac{\det \bigl(\Do^{\,\prime}e^{-K(x)}\gamma \Ko^{\,\prime}\bigr)}{
\displaystyle\prod_{1\leq {j}<j^{\prime }\leq {N_{a}}}(a_{j}-a_{j^{\prime }})
},  \label{sol5.8}
\end{equation}
where in analogy with~(\ref{eAB}) we introduced the diagonal $(N_{a}+N_{b})\times ({N_{a}+N_{b})}$-matrix
\begin{equation}
e^{K(x)}=\diag\{e^{K_{n}(x)}\}_{n=1}^{N_{a}+N_{b}},  \label{eK}
\end{equation}
and the new constant $N_{b}\times (N_{a}+N_{b})$ and $(N_{a}+N_{b})\times {N_{a}}$ matrices
\begin{align}
\Ko& =\Vert \Ko_{ln}\Vert , & \Ko_{ln}& =\prod_{l^{\prime }=1,\,l^{\prime
}\neq l}^{N_{b}}(\kappa _{n}-b_{l^{\prime }}),\quad l=1,\ldots ,N_{b},\quad
n=1,\ldots ,N_{a}+N_{b},  \label{152} \\
\Ko^{\,\prime}& =\Vert \Ko_{nj}^{\prime }\Vert , & \Ko_{nj}^{\prime }&
=\prod_{j^{\prime }=1,\,j^{\prime }\neq j}^{N_{a}}(a_{j^{\prime }}-\kappa
_{n}),\quad n=1,\ldots ,N_{a}+N_{b},\quad j=1,\ldots ,N_{a}.  \label{151}
\end{align}
Notice that they have a block structure since $\Ko_{l,N_{a}+l^{\prime }}$ and $\Ko_{j^{\prime }j}^{\,\prime }$ ($l,l^{\prime }=1,\dots ,N_{b}$,\thinspace $j,j^{\prime }=1,\dots ,N_{a}$) are diagonal submatrices. We also emphasize that, since $N_{a},N_{b}\geq 1$, as stated in (\ref{Nab}), the constant matrices $\Ko$, $\Ko^{\,\prime}$, $\Do$ and $\Do^{\,\prime}$ are not squared and, therefore, the determinants cannot be decomposed into products of determinants, which would imply $u(x)\equiv 0$.

In order to prove~(\ref{sol5.8}), let us notice that
\begin{align*}
& \bigl(\Ko e^{K(x)}\Do\bigr)_{l,l^{\prime }=1}^{N_{b}}\equiv \left(
\sum_{n=1}^{N_{a}+N_{b}}\left( \prod_{l^{\prime \prime }=1,\,l^{\prime
\prime }\neq l}^{N_{b}}(\kappa _{n}-b_{l^{\prime \prime }})\right)
e^{K_{n}(x)}\Do_{nl^{\prime }}\right) _{l,l^{\prime }=1}^{N_{b}} \\
& \quad=\sum_{j=1}^{N_{a}}e^{A_{j}(x)}d_{jl^{\prime }}\prod_{l^{\prime \prime
}=1,\,l^{\prime \prime }\neq l}^{N_{b}}(a_{j}-b_{l^{\prime \prime
}})+\sum_{l^{\prime \prime \prime }=1}^{N_{b}}e^{B_{l^{\prime \prime \prime
}}(x)}\delta _{l^{\prime \prime \prime },l^{\prime }}\prod_{l^{\prime \prime
}=1,\,l^{\prime \prime }\neq l}^{N_{b}}(b_{l^{\prime \prime \prime
}}-b_{l^{\prime \prime }}),
\end{align*}
where~(\ref{block}),~(\ref{eK}) and~(\ref{152}) were used. All terms in the last sum vanish except for $l^{\prime \prime \prime }=l$. Thus by~(\ref{sol13:2}) we get
\begin{align*}
& \det \bigl(\Ko e^{K(x)}\Do\bigr)_{l,l^{\prime }=1}^{N_{b}} \\
&\quad =(-1)^{N_{b}(N_{b}-1)/2}\left( \prod_{1\leq {l}<l^{\prime }\leq {N_{b}}
}(b_{l^{\prime }}-b_{l})^{2}\right) \det \left( \delta _{ll^{\prime \prime
}}e^{B_{l}(x)}+\sum_{j=1}^{N_{a}}\frac{e^{A_{j}(x)}c_{jl^{\prime }}}{
a_{j}-b_{l}}\right) _{l,l^{\prime }=1}^{N_{b}}.
\end{align*}
The determinant in the rhs coincides with the determinant in~(\ref{sol12:2}), which proves the first equality in~(\ref{sol5.8}). The second one is reduced to~(\ref{sol12:1}) by analogy, where also the explicit expression for matrix $\gamma $ in~(\ref{d11:1}) must be taken into account.

By considering convenient linear operations on the rows and columns of the matrices in~(\ref{sol5.8}) one can get more symmetric expressions of $\tau(x)$ and $\tau ^{\prime }(x)$ which involve only $\kappa $'s. Let us consider the first equality in~(\ref{sol5.8}). We can subtract the last row of matrix $\Ko$ from all the preceding rows. Then the $l$-th ($l>1$) row will read as $(b_{l}-b_{N_{b}})\prod_{l^{\prime }=1,\,l^{\prime }\neq l}^{N_{b}-1}(\kappa _{n}-b_{l^{\prime }})$. Each factor $b_{l}-b_{N_{b}}$ for $l=1,\dots ,N_{b}$ can then be extracted from the determinant and we repeat the same procedure with the last but one row, and up to the second one. The first row will then have all $1$'s, and the second row will have entries $\kappa _{l}-b_{1}$ for $l=1,\dots ,N_{a}+N_{b}$. Then one can easily shift $\kappa _{l}-b_{1}\rightarrow \kappa _{l}$ by using the first row. A similar transformation can be used to transform the generic element of the subsequent rows into $\kappa _{n}^{j}$. Finally, instead of~(\ref{sol5.8}) one obtains
\begin{equation}
\tau (x)=\det \left( \Vo e^{K(x)}\Do\right) ,\qquad \tau ^{\prime }(x)=\det
\left( \Do^{\,\prime}e^{-K(x)}\gamma \Vo^{\,\prime}\right) ,  \label{tau175}
\end{equation}
where by $\Vo$ and $\Vo^{\,\prime}$ we denote the \textquotedblleft incomplete Vandermonde matrices,\textquotedblright\ i.e., the $N_{b}\times (N_{a}+N_{b})$- and $(N_{a}+N_{b})\times {N_{a}}$-matrices given as
\begin{equation}
\Vo=\left(
\begin{array}{lll}
1 & \ldots  & 1 \\
\vdots  &  & \vdots  \\
\kappa _{1}^{N_{b}-1} & \ldots  & \kappa _{N_{a}+N_{b}}^{N_{b}-1}
\end{array}
\right) ,\qquad \Vo^{\,\prime}=\left(
\begin{array}{lll}
1 & \ldots  & \kappa _{1}^{N_{a}-1} \\
\vdots  &  & \vdots  \\
1 & \ldots  & \kappa _{N_{a}+N_{b}}^{N_{a}-1}
\end{array}
\right) .  \label{sol18}
\end{equation}

\begin{remark}
In the expressions~(\ref{sol5.8}) all objects with the exception of $K_{n}(x)$ are invariant with respect to an overall shift of all parameters $a$'s and $b$'s (or, equivalently, all $\kappa $'s) by the same constant, while in~(\ref{sol18}) this invariance is not obvious. In fact, following the same procedure used for transforming~(\ref{sol5.8}) into~(\ref{sol18}) one can get matrices $\Vo$ and $\Vo^{\,\prime}$ constructed from powers of $\kappa _{n}+z$ instead of $\kappa _{n}$, where $z$ is totally arbitrary.
\end{remark}

The expressions for the potential are invariant with respect to rescaling~(\ref{sol10}) and~(\ref{sol15}) and, consequently, we have by~(\ref{starting})
\begin{equation}
u(x)=-2\partial _{x_{1}}^{2}\log \tau (x)=-2\partial _{x_{1}}^{2}\log \tau
^{\prime }(x).  \label{sol19:0}
\end{equation}
This twofold expression for the potential follows also directly, as thanks to~(\ref{sol10}),~(\ref{sol12}),~(\ref{An}) and~(\ref{sol15}),
\begin{equation}
\tau (x)=(-1)^{N_{a}N_{b}+N_{a}(N_{a}-1)/2}\left(
\prod_{n=1}^{N_{a}+N_{b}}e^{K_{n}(x)}\right) V(\kappa _{1},\ldots ,\kappa
_{N_{a}+N_{b}})\tau ^{\prime }(x),  \label{tautau}
\end{equation}
where $V$ denotes the Vandermonde determinant
\begin{align}
V(\kappa _{1},\ldots ,\kappa _{N_{a}+N_{b}})& =\det \left(
\begin{array}{lll}
1 & \ldots  & 1 \\
\vdots  &  & \vdots  \\
\kappa _{1}^{N_{a}+N_{b}-1} & \ldots  & \kappa _{N_{a}+N_{b}}^{N_{a}+N_{b}-1}
\end{array}
\right)  \notag \\
& \equiv \prod_{1\leq m<n\leq {N_{a}+N_{b}}}(\kappa _{n}-\kappa _{m}).
\label{sol20:1}
\end{align}

Thus, we have shown the equivalence of representations~(\ref{starting}) and~(\ref{solex3:1}) for the potential with those expressed as determinants of product of three and four matrices given in~(\ref{tau175}). These representations coincide with the representations obtained using the $\tau $-function approach in~\cite{BC} and studied in detail in~\cite{BC2,B07,ChK1,ChK2}. Notice that the special block form of the matrices $\Do$ and $\Do^{\,\prime}$ in~(\ref{block}), in general, is not preserved, when in the determinants~(\ref{tau175}) we perform a renumbering of the parameters $\kappa _{n}$, for instance in order to have $\kappa _{1}<\kappa _{2}<\dots <\kappa _{N_{a}+N_{b}}$. This problem will be studied in details in section~\ref{invariance}.

\subsection{Explicit representation for the tau-functions}

In order to study the behaviour of the potential and Jost solutions, which we plan to perform in a forthcoming publication, it is convenient to derive an explicit representations for the determinants involved, see also~\cite{BK,BC,BC2,B07,ChK1,ChK2}. These representations involve the maximal minors of matrices $\Do$ and $\Do^{\,\prime}$ (see~(\ref{block})) for which we use the simplified notations:
\begin{equation}
\Do(n_{1},\ldots ,n_{N_{b}})=\Do\left(
\begin{array}{llll}
n_{1}, & n_{2}, & \ldots , & n_{N_{b}} \\
1, & 2, & \ldots , & N_{b}
\end{array}
\right) ,  \label{sol20:32}
\end{equation}
i.e., determinant of the $N_{b}\times {N_{b}}$-matrix, that consists of $\{n_{1},\ldots ,n_{N_{b}}\}$ rows of the matrix $\Do$ and all its columns, and
\begin{equation}
\Do^{\,\prime}(n_{1},\ldots ,n_{N_{a}})=\Do^{\,\prime}\left(
\begin{array}{llll}
1, & 2, & \ldots , & N_{a} \\
n_{1}, & n_{2}, & \ldots , & n_{N_{a}}
\end{array}
\right) ,  \label{sol20:31}
\end{equation}
i.e., determinant of the $N_{a}\times {N_{a}}$-matrix that consists of all rows of the matrix $\Do^{\,\prime}$ and $\{n_{1},\ldots ,n_{N_{a}}\}$ columns of this matrix. Then, by using the Binet--Cauchy formula for the determinant of a product of matrices and notation~(\ref{sol20:1}), we can rewrite relations~(\ref{tau175}) in the form
\begin{subequations}
\label{sol25.1}
\begin{align}
& \tau (x)=\sum_{1\leq n_{1}<n_{2}<\cdots <n_{N_{b}}\leq {N_{a}+N_{b}}
}f_{n_{1},\ldots ,n_{N_{b}}}\prod_{l=1}^{N_{b}}e^{K_{n_{l}}(x)},
\label{sol252} \\
& \tau ^{\prime }(x)=\sum_{1\leq n_{1}<n_{2}<\cdots <n_{N_{a}}\leq {
N_{a}+N_{b}}}f_{n_{1},\ldots ,n_{N_{a}}}^{\prime
}\prod_{j=1}^{N_{a}}e^{-K_{n_{j}}(x)},  \label{sol251}
\end{align}
\end{subequations}
where
\begin{subequations}
\label{sol23.1}
\begin{align}
f_{n_{1},n_{2},\ldots ,n_{N_{b}}}& =V(\kappa _{n_{1}}^{{}},\ldots ,\kappa
_{n_{N_{b}}}^{{}})\Do(n_{1},\ldots ,n_{N_{b}}),  \label{sol23:2} \\
f_{n_{1},n_{2},\ldots ,n_{N_{a}}}^{\prime }& =V(\kappa _{n_{1}}^{{}},\ldots
,\kappa _{n_{N_{a}}}^{{}})\Do^{\,\prime}(n_{1},\ldots
,n_{N_{a}})\prod_{j=1}^{N_{a}}\gamma _{n_{j}}.  \label{sol23:1}
\end{align}
\end{subequations}
From~(\ref{tautau}) it follows that the coefficients $f$ and $f^{\prime}$ in the two expansions~(\ref{sol251}) and~(\ref{sol252}) are related by the equation
\begin{equation}
f_{n_{1},\ldots ,n_{N_{a}}}^{\prime }=(-1)^{N_{a}N_{b}+N_{a}(N_{a}-1)/2}
\frac{f_{\widetilde{n}_{1},\widetilde{n}_{2},\ldots ,\widetilde{n}_{N_{b}}}}{
V(\kappa _{1},\ldots ,\kappa _{N_{a}+N_{b}})},  \label{twofrelated}
\end{equation}
where $\{n_{1},\ldots ,n_{N_{a}}\}$ and $\{\widetilde{n}_{1},\widetilde{n}_{2},\ldots ,\widetilde{n}_{N_{b}}\}$ are two disjoint ordered subsets of the set of numbers running from $1$ to $N_{a}+N_{b}$.

Let us mention that in our construction the two equivalent representations for the $\tau $-functions in~(\ref{tautau}) were obtained as a consequence of the two equalities in~(\ref{asymptk1}), while, of course, they can be proved directly by means of~(\ref{sol25.1}) and~(\ref{twofrelated}). See~\cite{ChK1}, where this property is called duality.

\begin{remark}
Notice that, in agreement with (\ref{sol23.1}) and (\ref{twofrelated}), we have
\begin{equation}
\prod\limits_{j=1}^{N_{a}}\gamma _{n_{j}}V(\kappa _{n_{1}},\dots ,\kappa
_{n_{N_{a}}})=\det \pi (-1)^{N_{a}N_{b}+N_{a}(N_{a}-1)/2}\frac{V(\kappa _{
\widetilde{n}_{1}},\dots ,\kappa _{\widetilde{n}_{N_{b}}})}{V(\kappa
_{1},\dots ,\kappa _{N_{a}+N_{b}})}.
\end{equation}
and
\begin{equation}
\mathcal{D}^{\prime }(n_{1},\dots ,n_{N_{a}})=\det \pi \mathcal{D}(
\widetilde{n}_{1},\dots ,\widetilde{n}_{N_{b}}),
\end{equation}
where $\pi $ is the matrix performing the permutation from $(\kappa_{n_{1}},\dots ,\kappa _{n_{N_{a}}},\kappa _{\widetilde{n}_{1}},\dots,\kappa _{\widetilde{n}_{N_{b}}})$ to $(\kappa _{1},\dots ,\kappa_{N_{a}+N_{b}})$.
\end{remark}

\section{Jost solutions and invariance properties}

\subsection{Properties of matrices $\Do$ and $\Do^{\,\prime}$}

Matrices $\Do$ and $\Do^{\,\prime}$ introduced in~(\ref{block}) obey rather interesting properties. As follows directly from the definition, they are orthogonal in the sense that
\begin{equation}
\Do^{\,\prime}\Do=0,  \label{d12}
\end{equation}
where zero in the rhs is a $N_{a}\times {N_{b}}$-matrix. Moreover, since the matrices
\begin{equation}
{\Do}^{\dag }\Do=E_{N_{b}}+d^{\dag }d,\qquad \Do^{\,\prime}{\Do^{\,\prime}}
^{\dag }=E_{N_{a}}+dd^{\dag },  \label{d14}
\end{equation}
where $\dag $ denotes Hermitian conjugation of matrices (in fact, transposition here), are invertible, the matrices (see~\cite{G1990})
\begin{align}
& \bigl(\Do\bigr)^{(-1)}=(E_{N_{b}}+d^{\dag }d)^{-1}{\Do}^{\dag },
\label{d16} \\
& \bigl(\Do^{\,\prime}\bigr)^{(-1)}={\Do^{\,\prime}}^{\dag
}(E_{N_{a}}+dd^{\dag })^{-1},  \label{d15}
\end{align}
are, respectively, the left inverse of the matrix $\Do$ and the right inverse of the matrix $\Do^{\,\prime}$, i.e.,
\begin{equation}
\bigl(\Do\bigr)^{(-1)}\Do=E_{N_{b}},\qquad \Do^{\,\prime}\bigl(\Do^{\,\prime}
\bigr)^{(-1)}=E_{N_{a}}.  \label{d17}
\end{equation}
Products of these matrices in the opposite order give the real self-adjoint $(N_{a}+N_{b})\times (N_{a}+N_{b})$-matrices
\begin{align}
& P=\Do\bigl(\Do\bigr)^{(-1)}=\Do(E_{N_{b}}+d^{\dag }d)^{-1}\bigl(\Do\bigr)
^{\dag },  \label{d19} \\
& P^{\,\prime }=\bigl(\Do^{\,\prime}\bigr)^{(-1)}\Do^{\,\prime}=\bigl(\Do
^{\,\prime }\bigr)^{\dag }(E_{N_{a}}+dd^{\dag })^{-1}\Do^{\,\prime},
\label{d18}
\end{align}
which are orthogonal projectors, i.e.,
\begin{equation}
P^{2}=P,\qquad (P^{\,\prime })^{2}=P^{\,\prime },\qquad PP^{\,\prime }=0=P^{\,\prime
}P,  \label{d22}
\end{equation}
and complementary in the sense that
\begin{equation}
P+P^{\,\prime }=E_{N_{a}+N_{b}}.  \label{d23}
\end{equation}
Orthogonality of the projectors follows from~(\ref{d12}) and the last equality from obvious relations of the kind $(E_{N_{b}}+d^{\dag }d)^{-1}d^{\dag}=d^{\dag }(E_{N_{a}}+dd^{\dag })^{-1}$.

\subsection{Symmetric representations for the Jost solutions}

In order to get a $\tau $-representation for the Jost solutions, we use~(\ref{sym34}) and notice that rescaling~(\ref{sol15}) does not modify the substitution rules given in~(\ref{sol11}). In terms of notations~(\ref{sol12}), (\ref{An}) and~(\ref{eK}) these rules read as
\begin{subequations}
\label{sol110}
\begin{align}
\tau (x)&\rightarrow \tau _{\chi }(x,\bk)\quad  \text{replacing}\quad
e^{K}\rightarrow e^{K}(\kappa +i\bk),  \label{sol110:2} \\
\tau ^{\prime }(x)&\rightarrow \tau _{\chi }^{\prime }(x,\bk)\quad\text{replacing}\quad e^{K}\rightarrow e^{K}(\kappa +i\bk),  \label{sol110:1} \\
\tau (x)&\rightarrow \tau _{\xi }(x,\bk) \quad\text{replacing}\quad
e^{K}\rightarrow \frac{e^{K}}{\kappa +i\bk},  \label{sol110:4} \\
\tau ^{\prime }(x)&\rightarrow \tau _{\xi }^{\prime }(x,\bk)\quad\text{
replacing}\quad e^{K}\rightarrow \frac{e^{K}}{\kappa +i\bk},
\label{sol110:3}
\end{align}
\end{subequations}
where $\kappa +i\bk$ denotes the diagonal $(N_{a}+N_{b})\times (N_{a}+N_{b})$-matrix
\begin{equation}
\kappa +i\bk=\diag\{\kappa _{1}^{{}}+i\bk,\ldots ,\kappa
_{N_{a}+N_{b}}^{{}}+i\bk\}  \label{diagk}
\end{equation}
and analogously for the matrix $(\kappa +i\bk)^{-1}$. Explicitly these replacements give (see~(\ref{tau175}))
\begin{subequations}
\label{tauk}
\begin{align}
\tau _{\chi }(x,\bk)& =\det \left( \Vo e^{K(x)}(\kappa +i\bk)\Do\right) ,
\label{sol172} \\
\tau _{\chi }^{\prime }(x,\bk)& =\det \left( \Do^{\,\prime}e^{-K(x)}(\kappa +i
\bk)^{-1}\gamma \Vo^{\,\prime}\right) ,  \label{sol171} \\
\tau _{\xi }(x,\bk)& =\det \left( \Vo e^{K(x)}(\kappa +i\bk)^{-1}\Do\right) ,
\label{sol174} \\
\tau _{\xi }^{\prime }(x,\bk)& =\det \left( \Do^{\,\prime}e^{-K(x)}(\kappa +i
\bk)\gamma \Vo^{\,\prime}\right) ,  \label{sol173}
\end{align}
\end{subequations}
Thus, by~(\ref{sym34}) we have
\begin{subequations}
\label{sym340}
\begin{align}
& \left( \prod_{l=1}^{N_{b}}(b_{l}+i\bk)\right) \chi (x,\bk)=\frac{\tau
_{\chi }(x,\bk)}{\tau (x)}\equiv \left( \prod_{n=1}^{N_{a}+N_{b}}(\kappa
_{n}+i\bk)\right) \frac{\tau _{\chi }^{\prime }(x,\bk)}{\tau ^{\prime }(x)},
\label{sym3:10} \\
& \left( \prod_{l=1}^{N_{b}}(b_{l}+i\bk)^{-1}\right) \xi (x,\bk)=\frac{\tau
_{\xi }(x,\bk)}{\tau (x)}\equiv \left( \prod_{n=1}^{N_{a}+N_{b}}(\kappa
_{n}+i\bk)^{-1}\right) \frac{\tau _{\xi }^{\prime }(x,\bk)}{\tau ^{\prime
}(x)}.  \label{sym4:10}
\end{align}
\end{subequations}
The Jost solutions themselves are then given by~(\ref{chiPhi}) and in order to simplify their analyticity properties it is convenient to renormalize them in the following way
\begin{equation}
\Phi (x,\bk)\rightarrow \frac{\Phi (x,\bk)}{\displaystyle
\prod_{l=1}^{N_{b}}(b_{l}+i\bk)},\qquad \Psi (x,\bk)\rightarrow \left(
\prod_{l=1}^{N_{b}}(b_{l}+i\bk)\right) \Psi (x,\bk).
\end{equation}
Then, to preserve relations given in~(\ref{chiPhi}) we also normalize $\chi(x,\bk)$ and $\xi (x,\bk)$ accordingly, so that instead of~(\ref{sym340}) we have
\begin{subequations}
\label{sym341}
\begin{align}
\chi (x,\bk)& =\frac{\tau _{\chi }(x,\bk)}{\tau (x)}\equiv \left(
\prod_{n=1}^{N_{a}+N_{b}}(\kappa _{n}+i\bk)\right) \frac{\tau _{\chi
}^{\prime }(x,\bk)}{\tau ^{\prime }(x)},  \label{sym3:11} \\
\xi (x,\bk)& =\frac{\tau _{\xi }(x,\bk)}{\tau (x)}\equiv \left(
\prod_{n=1}^{N_{a}+N_{b}}(\kappa _{n}+i\bk)^{-1}\right) \frac{\tau _{\xi
}^{\prime }(x,\bk)}{\tau ^{\prime }(x)}.  \label{sym4:11}
\end{align}
\end{subequations}
Now $\chi (x,\bk)$ is a polynomial with respect to $\bk$ of the order $\bk^{N_{b}}$ and $\xi (x,\bk)$ is a meromorphic function of $\bk$ that becomes a polynomial of the order $\bk^{N_{a}}$ after multiplication by $\prod_{n=1}^{N_{a}+N_{b}}(\kappa _{n}+i\bk)$. In other words, now $\Phi (x,\bk)$ is an entire function of $\bk$ and $\Psi (x,\bk)$ is meromorphic with poles at all points $\bk=i\kappa _{n}$, $n=1,\ldots ,N_{a}+N_{b}$. Introducing the discrete values of $\Phi (x,\bk)$ at these points as a $(N_{a}+N_{b})$-row
\begin{equation}
\Phi (x,i\kappa )=\{\Phi (x,i\kappa _{1}^{{}}),\ldots ,\Phi (x,i\kappa
_{N_{a}+N_{b}}^{{}})\},  \label{d6:11}
\end{equation}
and the residuals of $\Psi (x,\bk)$ at these points
\begin{equation}
\Psi _{\kappa _{n}}(x)=\res_{\bk=i\kappa _{n}}\Psi (x,\bk),  \label{resn}
\end{equation}
as a $(N_{a}+N_{b})$-column
\begin{equation}
\Psi _{\kappa }(x)=\{\Psi _{\kappa _{1}}(x),\ldots ,\Psi _{\kappa
_{N_{a}+N_{b}}}(x)\}^{\text{T}},  \label{d6:12}
\end{equation}
we get thanks to~(\ref{sol12}),~(\ref{sol13:2}) and~(\ref{block}) that the relations~(\ref{PhiPsi}) take the more symmetric form
\begin{equation}
\Phi (x,i\kappa )\Do=0,\qquad \Do^{\,\prime}\Psi _{\kappa }(x)=0.  \label{d6}
\end{equation}
It is necessary to mention that after the renormalization~(\ref{sym340}) the asymptotic conditions~(\ref{asymptk}) become
\begin{equation}
\lim_{\bk\rightarrow \infty }(i\bk)^{-N_{b}}\chi (x,\bk)=1,\qquad \lim_{\bk
\rightarrow \infty }(i\bk)^{N_{b}}\xi (x,\bk)=1,  \label{asymptk2}
\end{equation}
and the relations~(\ref{asymptk1}) take the form
\begin{equation}
u(x)=-2\lim_{\bk\rightarrow \infty }(i\bk)^{-N_{b}+1}\partial _{x_{1}}\chi
(x,\bk)=2\lim_{\bk\rightarrow \infty }(i\bk)^{N_{b}+1}\partial _{x_{1}}\xi
(x,\bk).  \label{asymptk3}
\end{equation}

Finally let us point out that by introducing an infinite set of times, and, precisely, replacing $K_{n}(x)$ with the formal series (see \cite{Miwa})
\begin{equation}
K_{n}(t_{1},t_{2},\ldots )=\sum_{j=1}^{\infty}\kappa_{n}^{j}t_{j},
\label{Kn}
\end{equation}
appropriate choices of the times $t_{j}$'s provide the multisoliton solutions of any nonlinear evolution equation in the hierarchy related to the KPII equation, as well as the corresponding Jost solutions. In particular the multisoliton solutions of the KPII equation are obtained by choosing $t_{1}=x_{1}$, $t_{2}=x_{2}$, $t_{3}=-4t$ and all highest times equal to zero.

\begin{remark}
\label{Miwa} We showed in ~(\ref{sol110}) that the Jost solutions can be obtained by means of transformations, that are equivalent to the formal Miwa shift used in the construction of the Baker--Akhiezer functions in terms of the $\tau $-functions. In fact, substitutions~(\ref{sol110:1}) and~(\ref{sol110:2}) can be obtained (up to an unessential factor $i\bk$) by considering the multi-time $\tau $-function obtained by replacing $K_{n}$ with the infinite formal series in~(\ref{Kn}) and, then, by shifting $t_{j}\rightarrow{t_{j}-1/j(i/\bk)^{j}}$. Similarly, the substitutions~(\ref{sol110:3}) and~(\ref{sol110:4}) are obtained (again up to an unessential factor $-i/
\bk$) by the shifts $t_{j}\rightarrow {t_{j}+1/j(i/\bk)^{j}}$. Then, non formal Jost solutions of the heat equation with a potential being a solution of KPII are derived by choosing $t_{1}=x_{1}$, $t_{2}=x_{2}$, $t_{3}=-4t$ and all highest times equal to zero.
\end{remark}

\subsection{Invariance properties of the multisoliton potential}

\label{invariance}In some cases it is useful to rename the spectral parameters $\kappa_{n}\rightarrow \widetilde{\kappa}_{n}$, for instance in such a way that the renamed parameters $\widetilde{\kappa}_{n}$ are ordered as follows
\begin{equation}
\widetilde{\kappa}_{1}<\widetilde{\kappa}_{2}<\dots <\widetilde{\kappa}_{
\mathcal{N}}.  \label{perm}
\end{equation}
This permutation can be performed by means of a $(N_{a}+N_{b})
\times(N_{a}+N_{b})$-matrix $\pi $ such that
\begin{equation}
(\widetilde{\kappa}_{1}^{},\ldots ,\widetilde{\kappa}_{N_{a}+N_{b}}^{})=(
\kappa_{1}^{},\ldots , \kappa_{N_{a}+N_{b}}^{})\pi .  \label{pi}
\end{equation}
This matrix is unitary,
\begin{equation}
\pi ^{\dag}=\pi ^{-1},  \label{d2:1}
\end{equation}
and all rows and columns have one element equal to 1 and all other equal to 0. It is convenient to write this matrix in a block form like
\begin{equation}
\pi =\left(
\begin{array}{ll}
\pi_{11} & \pi_{12} \\
\pi_{21} & \pi_{22}
\end{array}
\right) ,  \label{d3:2}
\end{equation}
where $\pi_{11}$ is a $N_{a}\times {N_{a}}$-matrix, $\pi_{22}$ a $N_{b}\times {N_{b}}$-matrix, $\pi_{12}$ a $N_{a}\times{N_{b}}$-matrix and $\pi_{21}$ a $N_{b}\times {N_{a}}$-matrix.

Then, representations~(\ref{tau175}) under the transformation~(\ref{pi}) keep the same form if the matrices $\Do$ and $\Do^{\,\prime}$ are transformed as follows
\begin{equation}
\Do\rightarrow \widetilde{\Do}=\pi \Do=\left(
\begin{array}{r}
d_{1} \\
d_{2}
\end{array}
\right) ,\qquad \Do^{\,\prime}\rightarrow \widetilde{\Do}^{\,\prime }=\Do
^{\,\prime }\pi ^{\dag }=(d_{1}^{\,\prime },-d_{2}^{\,\prime }),  \label{d2}
\end{equation}
where, from~(\ref{block}), we have
\begin{align}
d_{1}& =\pi _{11}d+\pi _{12},\qquad d_{2}=\pi _{21}d+\pi _{22}, \\
d_{1}^{\,\prime }& =\pi _{11}^{\dag }-d\pi _{12}^{\dag },\qquad -d_{2}^{\,\prime
}=\pi _{21}^{\dag }-d\pi _{22}^{\dag }.
\end{align}
While the size of blocks in~(\ref{d2}) are the same as in~(\ref{block}), the block structure in general is different. Nevertheless, thanks to the unitarity of the matrix $\pi $, relation~(\ref{d12}) remains valid for the transformed matrices, which by~(\ref{d2}) means that
\begin{equation}
d_{1}^{\,\prime }d_{1}=d_{2}^{\,\prime }d_{2}.  \label{d2:2}
\end{equation}
Also relations~(\ref{d6}) are preserved for the transformed quantities and transformed projectors $\widetilde{P}$ and $\widetilde{P}^{\,\prime }$ can be built.

Let us mention that the special block structure of the matrices $\Do$ and $\Do^{\,\prime}$ in~(\ref{block}) determines them uniquely in correspondence to a given potential. One can release this condition, without changing the potential, by multiplying the matrix $\Do$ by any nonsingular $N_{b}\times {N_{b}}$-matrix from the right and the matrix $\Do^{\,\prime}$ by any nonsingular $N_{a}\times {N_{a}}$-matrix from the left. In fact, determinants of these matrices cancel out in~(\ref{sol19:0}) and~(\ref{sym340}). Viceversa one can use this procedure for bringing $\widetilde{\Do}$ and $\widetilde{\Do}^{\,\prime }$ back from the block structure~(\ref{d2}) to the special block structure in~(\ref{block}). This can be done by using matrices $(d_{2})^{-1}$ and $(d_{1}^{\,\prime })^{-1}$, if they exist, to perform the following transformations
\begin{equation}
\widetilde{\Do}\rightarrow \widetilde{\Do}(d_{2})^{-1}=\left(
\begin{array}{l}
d_{1}(d_{2})^{-1} \\
E_{N_{b}}
\end{array}
\right) ,\qquad \widetilde{\Do}^{\,\prime }\rightarrow (d_{1}^{\,\prime })^{-1}
\widetilde{\Do}^{\,\prime }=(E_{N_{a}},-(d_{1}^{\,\prime })^{-1}d_{2}^{\,\prime }),
\label{d3}
\end{equation}
and, then, by noticing that thanks to~(\ref{d2:2})
\begin{equation}
d_{1}(d_{2})^{-1}=(d_{1}^{\,\prime })^{-1}d_{2}^{\,\prime }=\widetilde{d}.
\label{d3:1}
\end{equation}
Taking into account that both representations in~(\ref{tau175}) are equivalent, we deduce that matrices $d_{2}$ and $d_{1}^{\,\prime }$, if invertible, are simultaneously invertible. Thus, we proved that in the case of a permutation $\pi $ such that the matrix $d_{2}$ (or $d_{1}^{\,\prime }$) is nonsingular, the permutation of $\kappa $'s is equivalent to a transformation of the matrix $d$ to $\widetilde{d}$, or, correspondingly, by~(\ref{sol13:2}), to a transformation of the matrix $c$. This is always the case when the matrix $\pi $ has a diagonal block structure, i.e., when $\pi _{12}=\pi _{21}=0$. In this situation both matrices $d_{2}$ and $d_{1}^{\,\prime }$ are invertible and one can use the above substitution. Notice that in this case the permutation of $\kappa $'s does not mix the original parameters $a$'s with $b$'s (see (\ref{sol12})).

In general in making a permutation of $\kappa $'s, say, reducing them to the order given in~(\ref{perm}), the block structure~(\ref{block}) is lost and we can only say that the $\tau $-functions are given by~(\ref{tau175}) and~(\ref{tauk}), where both matrices $\Do$ and $\Do^{\,\prime}$ have at least two nonzero maximal minors.

\section{Concluding remarks}

\label{concl}Here we described relations between different representations, existing in the literature, of the multisoliton potentials of the heat operator and we derived forms of these representations that will enable, in a forthcoming publication, a detailed study of the asymptotic behaviour of the potentials themselves and the corresponding Jost solutions on the $x$-plane. We also presented various formulations of the conditions that guarantee the regularity of the potential. Nevertheless, the essential problems of determining the necessary conditions of regularity of the multisoliton potential is left open. In this context, let us mention the special interest of the specific subclass of potentials satisfying strict inequalities in~(\ref{condition}), which by~(\ref{sol252}) is equivalent to requiring that all $f_{n_{1}},\ldots ,_{N_{b}}$ are of the same sign (analogously, by~(\ref{sol251}) that all $f_{n_{1}}^{\prime },\ldots,_{N_{a}}$ are of the same signs). These conditions identify fully resonant soliton solutions (cf. \cite{BK,BC}). When such conditions are imposed, all maximal minors of matrices $\Do$ and $\Do^{\,\prime}$ are different from zero, as follows from~(\ref{sol23:2}) and~(\ref{sol23:1}) and, then, thanks to the invariance properties discussed above, we can always permute parameters $\kappa $'s in any way, for instance as in~(\ref{perm}), and at the same time deal with transformed matrices $\widetilde{\Do}$ and $\widetilde{\Do}^{\,\prime }$, which have a special block structure as in~(\ref{block}).

\section*{Acknowledgments}

This work is supported in part by the grant RFBR \# 08-01-00501, grant RFBR--CE \# 09-01-92433, Scientific Schools 795.2008.1, by the Program of RAS ``Mathematical Methods of the Nonlinear Dynamics,'' by INFN and by Consortium E.I.N.S.T.E.I.N. AKP thanks Department of Physics of the University of Salento (Lecce) for kind hospitality.

\end{document}